\documentclass[sigconf, nonacm]{acmart}

\AtBeginDocument{%
  }

\setcopyright{acmlicensed}
\copyrightyear{2026}
\acmYear{2018}
\acmDOI{XXXXXXX.XXXXXXX}
\acmConference[UIST'26]{Symposium on User Interface Software and Technology}{November 2--5,
  2026}{Detroit, Michigan}
\acmISBN{978-1-4503-XXXX-X/2026/11}

\usepackage{graphicx}
\usepackage{listings}
\usepackage{xcolor}

\begin{document}

\title{From Gaze to Guidance: Interpreting and Adapting to Users' Cognitive Needs with Multimodal Gaze-Aware AI Assistants}

\author{Valdemar Danry}
\orcid{0000-0001-5225-0077}
\affiliation{%
  \institution{Microsoft Research}
  \institution{MIT Media Lab}
  \city{Cambridge, MA}
  \country{USA}
}

\author{Javier Hernandez}
\orcid{0000-0001-9504-5217}
\affiliation{%
  \institution{Microsoft Research}
  \city{Cambridge, MA}
  \country{USA}
}

\author{Andrew Wilson}
\orcid{1234-5678-9012}
\affiliation{%
  \institution{Microsoft Research}
  \city{Cambridge}
  \state{Massachusetts}
  \country{USA}
}

\author{Pattie Maes}
\orcid{}
\affiliation{%
  \institution{MIT Media Lab}
  \city{Cambridge, MA}
  \country{USA}
}

\author{Judith Amores}
\orcid{0000-0003-1285-6909}
\affiliation{%
  \institution{Microsoft Research}
  \city{Cambridge, MA}
  \country{USA}
}

\renewcommand{\shortauthors}{Danry et al.}

\begin{abstract}
    Current LLM assistants are powerful at answering questions, but they have limited access to the behavioral context that reveals when and where a user is struggling. We present a gaze-grounded multimodal LLM assistant that uses egocentric video with gaze overlays to identify likely points of difficulty and target follow-up retrospective assistance. We instantiate this vision in a controlled study (n=36) comparing the gaze-aware AI assistant to a text-only LLM assistant. Compared to a conventional LLM assistant, the gaze-aware assistant was rated as significantly more accurate and personalized in its assessments of users’ reading behavior and significantly improved people's ability to recall information. Users spoke significantly fewer words with the gaze-aware assistant, indicating more efficient interactions. Qualitative results underscored both perceived benefits in comprehension and challenges when interpretations of gaze behaviors were inaccurate. Our findings suggest that gaze-aware LLM assistants can reason about cognitive needs to improve cognitive outcomes of users.
\end{abstract}

\ccsdesc[500]{Human-centered computing~Interactive systems and tools}

\keywords{AI agents, eye-gaze, reasoning}
\begin{teaserfigure}
  \includegraphics[width=\textwidth]{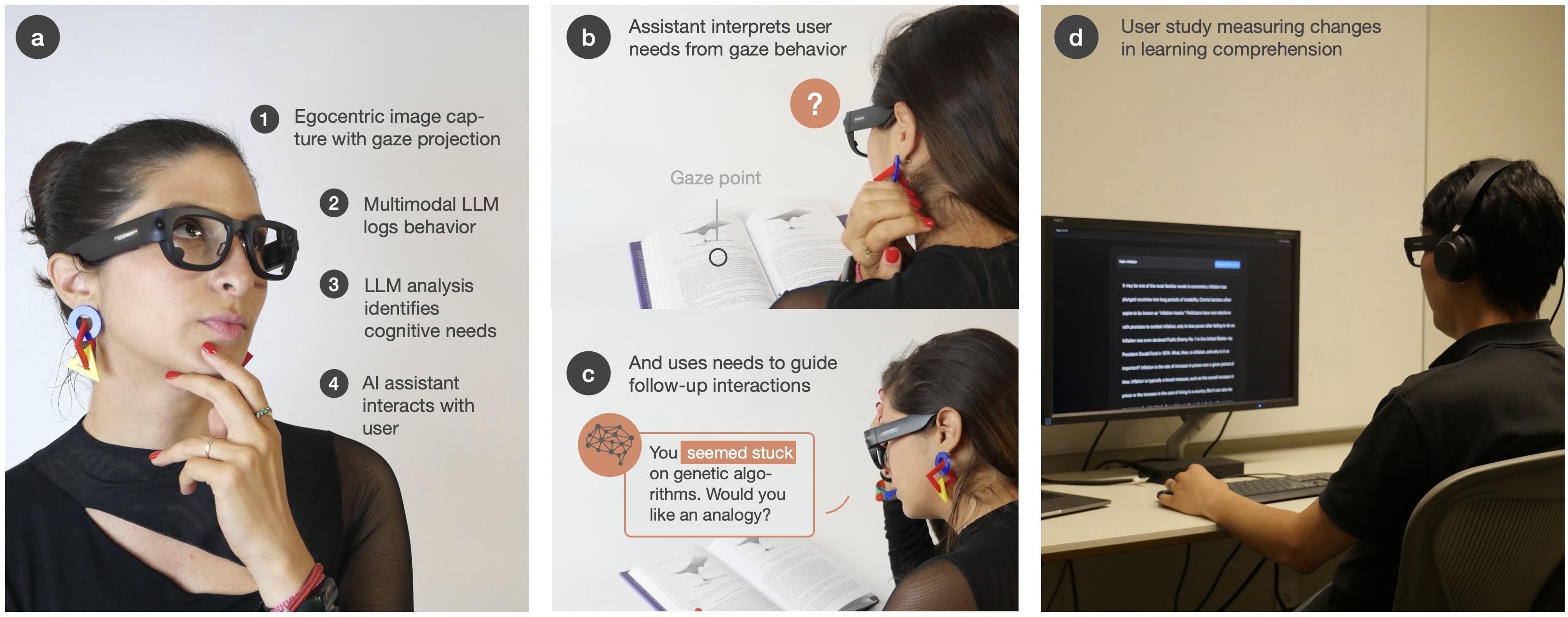}
  \caption{Overview of the gaze-aware cognitive AI assistant. (a)~A user wearing wearable AI glasses  which streams eye gaze data in to a gaze-aware cognitive LLM assistant. (b-c)~An example of the AI providing assistance after reading a difficult book, (d)~User study evaluating AI interaction and accuracy.}
  \Description{Four photographs illustrating the gaze-aware cognitive assistant prototype and study setup. 
(a) A participant wearing smart glasses looks thoughtful; text notes that physiological data such as egocentric video and eye tracking are captured, and the LLM assistant interprets needs. 
(b) The system identifies the user’s gaze point on a textbook while the assistant interprets needs from gaze behavior. 
(c) The assistant provides retrospective support, asking if the user would like an analogy after detecting difficulty with genetic algorithms. 
(d) Another participant, wearing glasses and headphones, sits at a computer during a user study measuring comprehension.}
  \label{fig:teaser}
\end{teaserfigure}


\maketitle

\section{Introduction}

In everyday life, humans routinely rely on other people's gaze to infer what they are focusing on, interested in, thinking about or struggling with. We follow others’ eyes to establish joint focus, infer the object of a referential expression, anticipate upcoming actions, assess if someone is actively thinking, and interpret social intentions such as interest, avoidance, confidence, or a desire to engage \cite{frischen2007gaze,canigueral2019role,johansson2001eye}. A student who repeatedly looks back and forth between the same lines of text may seem stuck; someone who looks at an object and then gazes upwards may be actively thinking about it; a person who averts their gaze or looks downward in conversation may appear hesitant or uncomfortable; a person staring dead into space may be tired or disinterested; and someone who lingers visually on an object or place may signal curiosity or interest \cite{johansson2001eye,canigueral2019role,argyle1994gaze}.

Being able to notice such moments matters because it changes how help can be given. Detecting when someone is confused, hesitant, disengaged, or curious allows others to respond with clarification, encouragement, reframing, or additional context at the moment it is most useful. In this sense, gaze is not only a cue to attention, but also a cue to possible cognitive need. When these moments are recognized and supported well, they can become opportunities for comprehension, persistence, and cognitive growth rather than silent failure or disengagement \cite{d2014confusion,stanton2021fostering,mirandola2018metacognitive}. This is especially important because people do not always reliably notice when their own understanding is breaking down, nor can they always articulate why they are stuck or what kind of help they need \cite{stanton2021fostering,mirandola2018metacognitive,marchand2007motivational}.

At the same time, today’s AI assistants and multimodal large language model (LLM) interfaces remain largely blind to this kind of behavioral context. Although modern assistants can understand text, images, audio, video, and support tasks such as question answering, summarization, and extended dialogue, they still rely primarily on explicit user input: what the user says, types, uploads, or asks. As a result, they are often unable to perceive the subtle behavioral signals that humans use to infer when help may be needed, what the user may be struggling with, or whether the user is confused, frustrated, disengaged, or curious before that state is explicitly verbalized.

In this paper, we explore this gap through the design and evaluation of a gaze-aware cognitive assistant that integrates egocentric sensing with multimodal LLM reasoning. Unlike prior efforts that use gaze metrics (such as fixation and regression analysis) to help the system know what the user is looking at or what the user is referring to, our system uses first-person video and projected gaze as direct multimodal context for an AI assistant, enabling the model itself to interpret behavior patterns to infer candidate indicators of cognitive need. We investigate whether grounding assistance in this behavioral context allows AI systems to offer more relevant, personalized, and cognitive-oriented support by shifting the interpretation of moments of difficulty from the user to the model, reducing the burden on users to explicitly recognize and articulate when and where they need help.

This paper makes the following contributions:

\begin{enumerate}
    
    \item A Gaze-Aware LLM assistant with a gaze-grounded prompting architecture that converts gaze-projected egocentric video into temporally structured observations for multimodal LLM assistance.
    
    \item A controlled study in a reading context showing improved recall and stronger perceived accuracy and targeting than a text-only baseline, alongside a characterization of failure modes.

    \item An approach and design implications for gaze-aware AI assistants that shifts gaze from a signal used primarily for attention tracking, reference grounding, or explicit control to gaze as a contextual behavioral input for multimodal LLM reasoning about user candidate moments of difficulty.
    
\end{enumerate}

\section{Related Work} 

\subsection{Gaze as a signal for adaptive assistance}

Eye tracking provides a rich but indirect signal about human cognition. Prior work has linked gaze behavior---including fixation duration, regressions, pupil dilation, and blink dynamics---to constructs such as effortful processing, cognitive load, mind wandering, and confusion, while social-gaze research has shown that gaze and gaze aversion can also reflect interactional states such as turn management, monitoring, and effort regulation \cite{hessels2024fundamentals,konig2016eye,gorin2024review,meziere2025eye,doherty2005gaze,abeles2017just}. Dynamic gaze visualizations have likewise been shown to support above-chance judgments of task, preference, answer choice, task performance, and confidence, especially when the gaze trace is shown in scene context \cite{van2018looking,bush2015viewing,emhardt2020inferring}. At the same time, gaze is fundamentally ambiguous: the same visual behavior may reflect confusion, verification, curiosity, familiarity, or planning, and is therefore best treated as probabilistic evidence rather than direct access to internal state \cite{hessels2024fundamentals,van2018looking,lim2020emotion}.

Building on this, prior systems have used gaze to detect possible cognitive needs and adapt support. \textit{GazeTutor} detected disengagement and triggered gaze-sensitive tutoring dialogue \cite{d2012gaze}. Other work has used gaze together with behavioral features and supervised models to predict confusion in learning environments and interactive visualizations \cite{pachman2016eye,lalle2016predicting}. Related gaze-contingent instructional systems adapt explanations or learning materials based on visual behavior, for example by providing personalized processing support in multimedia learning tasks \cite{scheiter2019adaptive}. More recent systems move from detection to assistance: \textit{The Reading Assistant} used gaze-triggered auditory prompting for reading support, while later work explored adaptive reading and writing assistance, including gaze-based augmentations for low-vision reading, situation-aware writing support, and mixed-reality reading help with LLMs \cite{sibert2000reading,wang2024gazeprompt,langner2023leveraging,thaqi2024sara}. Other recent systems combine gaze with LLM-generated feedback or GenAI adaptation in domains such as reading and programming \cite{santhosh2024gaze,schulz2025eyegpt}. However, these systems generally operate in text- or screen-centric settings and typically pass OCR text, detected text spans, or derived gaze metrics to the language model rather than using raw egocentric visual context as input.

\subsection{Gaze-grounded multimodal assistants}

A parallel line of work has explored gaze as an implicit multimodal cue for grounding interaction. In XR and embodied assistant settings, gaze has been used together with speech, pointing, actions, and dialogue history to disambiguate references, recover user intent, and localize relevant scene elements \cite{plopski2022eye,lee2024gazepointar,koelle2024embardiment,sarch2025grounding}. In these systems, gaze primarily helps determine \emph{what} the user refers to or \emph{which} object or subtask is relevant, rather than whether the user may be cognitively stuck on specific content.

Related work in egocentric assistance has studied proactive support from first-person video. \textit{HoloAssist} introduced a large-scale egocentric benchmark for real-world assistance, and later work synthesized dialogue and modeled proactive help from streaming egocentric video \cite{wang2023holoassist,zhang2025proactive}. Recent model-centric work has also begun incorporating gaze into first-person multimodal language model pipelines. For example, \textit{GazeLLM} uses gaze to guide efficient processing of first-person video, and \textit{EgoGazeVQA} studies whether gaze-guided prompting improves intent inference from egocentric daily-life video \cite{rekimoto2025gazellm,peng2025eye}. These systems demonstrate the feasibility of gaze-guided egocentric multimodal reasoning, but their emphasis is on referential grounding, efficiency, or task understanding rather than on targeting conversational support around likely moments of difficulty.

Taken together, prior work has used gaze mainly in three ways: for reference grounding in multimodal interaction, for OCR- or text-centric reading and writing support, or as a source of downstream metrics such as dwell time, regressions, or estimated cognitive load \cite{koelle2024embardiment,thaqi2024sara,langner2023leveraging,santhosh2024gaze,schulz2025eyegpt}. Our work differs in treating gaze-projected egocentric video as \emph{direct multimodal input} to the assistant and using temporal interpretation of gaze behavior over multiple timestamped images in scene context to surface candidate cognitive needs for support. Rather than using gaze only to resolve reference or only passing reduced gaze-derived features to the model,  we explore gaze as behavioral context for targeting assistance.

\subsection{Positioning our work}

Our work sits at the intersection of these threads but addresses a different gap. Compared with prior gaze-aware tutoring and reading systems, we do not limit assistance to text regions, OCR output, or predefined gaze metrics alone. Compared with gaze-grounded XR assistants, our goal is not only referential grounding but interpretation of gaze as a probabilistic cue to possible cognitive need. And compared with recent egocentric MLLM work, we are not primarily concerned with video understanding benchmarks, task replication, or compute-efficient cropping. Instead, we explore whether \emph{egocentric video with projected gaze overlays can serve as direct multimodal input to an interactive assistant that infers candidate moments of cognitive struggle and supports the user conversationally}. This shifts gaze from a narrow deictic or analytic feature into behavioral context for assistance, while also moving beyond stationary text-only settings toward more open-ended first-person interaction.

\begin{figure*}[ht]
    \centering
    \includegraphics[width=0.99\linewidth]{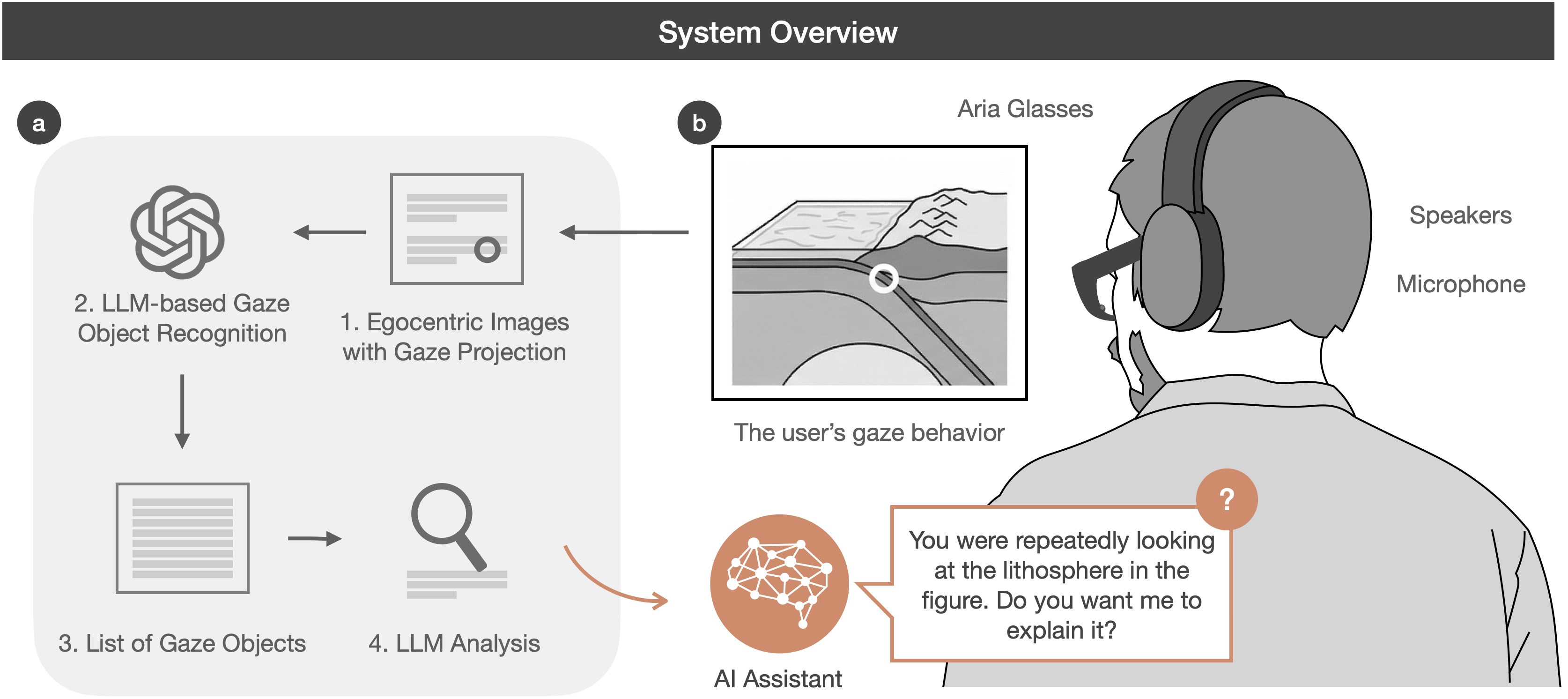}
    \caption{System overview of the gaze-aware AI assistant.}
    \label{fig:system-overview}
    \Description{Diagram of the gaze-aware AI assistant. 
(a) A pipeline shows how egocentric images with gaze points are processed through LLM-based gaze object recognition to generate a list of gaze objects, which are then analyzed by the LLM. 
(b) Illustration of a participant wearing Aria glasses with speakers and a microphone while reading a difficult text. 
The AI assistant detects repeated gaze on a word and asks, “You were repeatedly going back to this word. Do you want me to explain it?”}

\end{figure*}

\section{System Description}
\label{sec:implementation}



We designed a gaze-aware AI assistant that streams egocentric video and eye-tracking data from head-worn smart glasses to multimodal LLMs, which interpret the user's gaze behavior in context and provides after-the-fact conversational support (shown in Figure \ref{fig:system-overview}).

\subsection{Multimodal Scene Grounding}

The sensing setup consisted of the Meta Aria glasses streaming a forward-facing RGB point-of-view camera ($2880\times2880$) together with two inward-facing eye-tracking cameras. Gaze projection was computed locally using the open-source Aria gaze inference model\footnote{\url{https://github.com/facebookresearch/projectaria_eyetracking}}. The resulting gaze vector was projected on to the egocentric image as a red semi-transparent dot. 

A $200\times200$ pixel crop centered on the gaze point was generated in parallel, since previous work has shown that giving a multimodal language model a regular and zoomed frame increases accuracy \cite{zhang2023visual}. The gaze-projected images were given as direct input to the LLM. We chose this design instead of a conventional OCR-first (which is commonly used other approaches\cite{koelle2024embardiment, santhosh2024gaze}) to enable applications beyond reading; direct multimodal grounding preserves visual scene context around the attended region and makes the architecture more extensible beyond text-only settings, allowing the language model to make more contextually expansive interpretations of gaze behavior.

\subsection{Inference of Candidate Moments of Difficulty}

To avoid overwhelming the model \cite{zhang2024visual, khot2022decomposed}, the inference of candidate moments of difficulty was decomposed into two tasks: (1) creating a text-based list of gaze behavior over time, and (2) interpreting the list of behavior to infer candidate need states.

\begin{figure*}
    \centering
    \includegraphics[width=0.99\linewidth]{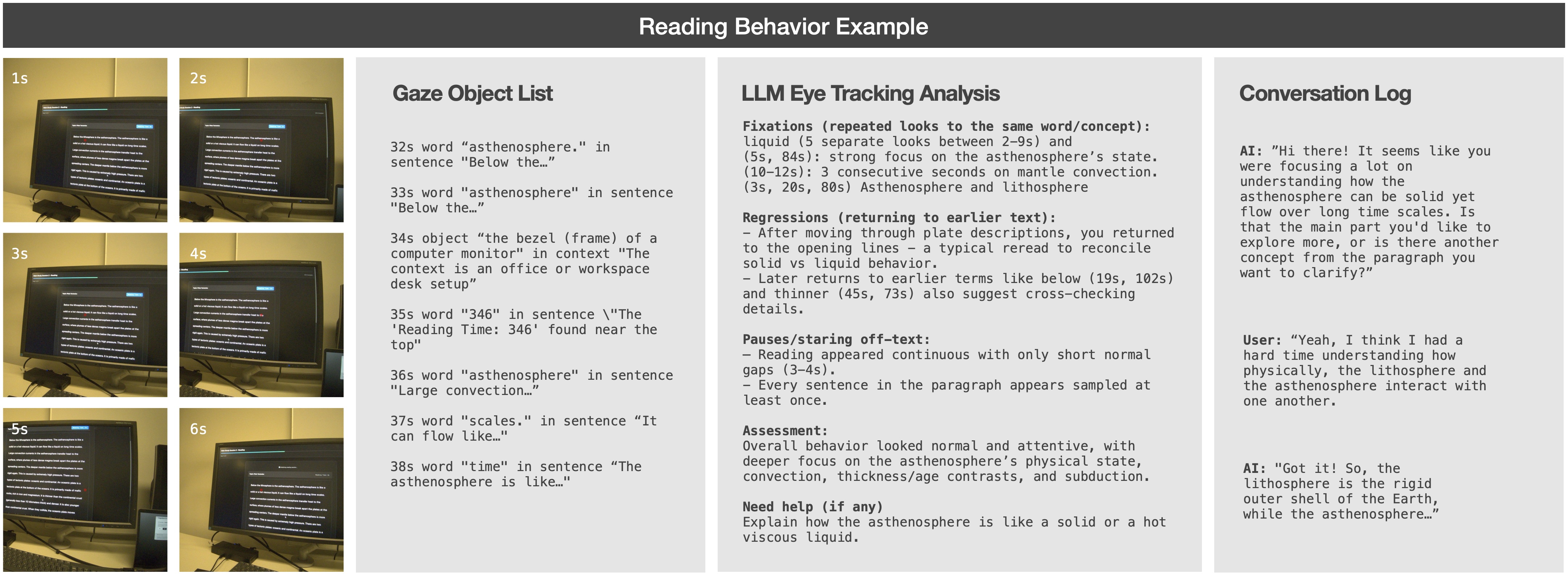}
    \caption{Real examples of gaze-reading behavior, the object list generated from LLM object recognition, the LLM eye tracking analysis with the object list as input, and a log of the resulting AI interaction.}
    \label{fig:reading-behavior-example}
    \Description{Composite figure showing how gaze data informs AI support. 
On the left, six sequential screenshots (1–6s) illustrate a participant’s gaze as they read text on a monitor. 
In the middle, a gaze object list logs words or objects fixated on, such as “asthenosphere,” with timestamps. 
Next, an LLM eye tracking analysis summarizes behaviors: fixations, regressions, off-text pauses, and an assessment noting difficulty with the asthenosphere’s physical state. 
On the right, a conversation log shows the AI asking if the user wants to clarify how the asthenosphere can be both solid and flowing, and the user confirming difficulty with lithosphere–asthenosphere interaction before the AI provides an explanation.}
\end{figure*}

For the first component, full egocentric frames with gaze overlay together with the local crop were sent to GPT-4.1 every 0.5 seconds. The model was then prompted to return a structured description of the attended item which was then timestamped and added to the action list:

\begin{verbatim}
class GazeAnalysis(BaseModel):
    type: Literal["word","object","none"]
    content: str   # e.g., "Quantum"
    context: str   # e.g., "In the sentence: 'Quantum is...'"
\end{verbatim}

For example, the model might return a word (e.g. 'Quantum') together with the sentence or phrase in which it appeared, or identify a non-text object (e.g. a lamp) and its context (e.g. standing on a desk). This structured representation gave us a simple intermediate output and helped to not overwhelm the model, which especially for multimodal inputs may struggle without task decomposition \cite{zhang2024visual, khot2022decomposed}. See an example of an action list in Fig. \ref{fig:reading-behavior-example}.

Next, the second component infers candidate need states from the gaze behavior action list. For this, the sequence of \texttt{GazeAnalysis} objects produced by the grounding stage was passed to GPT-4.1 prompted to identify gaze-linked behaviors and needs, and align them with specific words, phrases, or sentence regions (For the full prompt see Appendix \ref{appx:prompts}). The analyses results are then used to trigger or give context to the conversational assistant.



For the present study, this temporal analysis was performed retrospectively after the participant completed reading the paragraph. The grounding stage produced one timestamped observation every 0.5 seconds, and the full grounded gaze sequence for that trial was accumulated and processed only after reading ended. In other words, the analysis window in this evaluation was the complete reading episode for that passage, rather than a rolling online buffer. We adopted this boundary-aligned design after pilot testing because it avoided interrupting readers in the middle of a difficult passage, though it necessarily sacrificed immediacy of support. More generally, the same architecture could support alternative trigger policies in future work, including fixed-interval analysis, on-demand analysis when the user explicitly queries the assistant, or event-triggered just-in-time assistance.

\subsection{Conversational Assistance}

The third component of the system turns inferred candidate need states into assistance. The conversational layer was implemented using the OpenAI Realtime API (\texttt{gpt-4o-realtime-preview}) via Azure or OpenAI for low latency. The assistant received the text together with the gaze-based analysis produced by the previous stage (or, in the control condition, a text-only analysis without gaze). Audio was streamed as 24\,kHz PCM16 mono chunks, with speech recognition and synthesis handled by the same API. In typical use, round-trip speech latency remained below 2\,s, while the full pipeline from gaze-linked observation to available assistant output averaged 4.1\,s end-to-end. For the purpose of this study, the assistant was prompted to assist the user based on their cognitive needs and use both explanations and Socratic interactions \cite{danry2023don}. Moreover, given the uncertainty around interpreting cognitive needs from gaze alone, we prompted the assistant to use hedging and confirm with users before explaining as recommended in prior work \cite{hernandez2021guidelines}.

\subsection{Design Rationale}

Taken together, the architecture prioritizes (1) \emph{direct multimodal grounding} over OCR-only reduction in order to preserve scene context and generalize beyond text-only interfaces; (2) \emph{temporal reasoning problem} over frame-wise classification, since moments of need emerge from gaze patterns over time rather than isolated fixations; (3) \emph{selective interpretation} of gaze-linked events rather than indiscriminate ingestion of all available context, since more context is not always more useful for assistance; and (4) \emph{uncertainty-aware and low-friction}, using hedged language and conservative timing to reduce the risk of overclaiming and interruption.

We evaluate this architecture in reading comprehension because it offers a tractable first testbed: gaze can often be related to specific words, phrases, and passages, and the effects of support can be assessed in a more controlled way, using established comprehension measures. However, the system architecture itself is not inherently tied to reading. More broadly, it is intended as a prototype of how multimodal LLM assistants might use gaze-projected egocentric video to infer candidate cognitive need states and provide targeted support.

\section{Study Design}
To evaluate the impact of the AI assistant on user comprehension and experience, we conducted a controlled user study on a reading task. Our primary aim was to understand not only how gaze-aware cognitive assistants is able to accurately capture cognitive needs but also how it affects cognitive outcomes such as learning and how users interact with the assistant itself, compared to an assistant without access to egocentric gaze video.

\subsection{Participants and Data Collection}
We recruited 41 healthy adults~(age 18-44, largest group 25-34 years; gender 57\% male, 43\% female) across two independent study sites through various mailing lists. We excluded participants who did not speak English, had conditions preventing reading, or could not wear contact lenses in place of their own glasses. Five participants were excluded from analysis: three due to inadvertent device disconnection that caused data loss, and two for completing the study in under 10 minutes (average completion time was 51 minutes). All subjects gave written informed consent prior to participating (reviewed and approved by the Institutional Review Boards, and were given a \$50 AMEX gift card. All data was stored locally on a password-protected machine in accordance with the IRB in a de-identified manner. LLMs were accessed through the Azure AI Foundry via API endpoints and OpenAI's API. The study was partly completed at [REDACTED] location and completed at [REDACTED LOCATION].

\subsection{Experimental Conditions}
We conducted a within-subjects experimental design with two conditions: (1) an interactive assistant interpreting their cognitive needs from egocentric gaze video (gaze-aware cognitive assistant), and (2) an interactive assistant interpreting cognitive needs only from the text they read. In both conditions, users read one text while wearing the prototype headset with egocentric gaze streaming on or off. When the gaze streaming was on, the system compiled a gaze actions throughout the reading. After reading the text, users interacted with an assistant that either (1) had access to both the text they read and the LLM-generated analysis of cognitive needs from their gaze data (see Section \ref{sec:implementation}), or (2) had access to the text they read and text-based analysis with the exact prompting as the LLM-generated analysis of cognitive needs from their gaze data but without gaze data. For the prompts used in both conditions and examples of analyses see Appendix \ref{appx:prompts}. Participants communicated with the assistant through speech.

\subsection{Materials}
Six readings with topics of varying difficulty were selected from Wikipedia. For each topic, Flesch-Kincaid Grade (FKG) level was calculated to ensure variance in difficulty across selected texts. Word count (WC) was kept consistent to allow for comparison. Topics include (1)~"Water Cycle", (2)~"Plate tectonics", (3)~"Inflation", (4)~"Climate Change", (5)~"Superdeterminism", (6)~"Topological Quantum Computing".

To assess recall, transfer learning, and definitional learning, recall, concept-inventories and definition probes were constructed, following previous literature.
Recall questions were constructed following Roediger \& Karpicke's retrieval-practice questions~\cite{karpicke2008critical}. Each recall question consisted of three literal facts per passage. Definition probe questions were constructed to test the assistant's ability to notice terms undefined in the text that the user didn't understand. Lastly, to measure deeper understanding and transfer learning to new contexts by generalization, we constructed concept-inventory items, which in literature are used to evaluate transfer learning. Outcome metrics followed best practice from retrieval-practice research~\cite{karpicke2008critical}, incidental vocabulary assessment~\cite{nagy1985learning, nation2007vocabulary} and the concept-inventory methods originating with the Force Concept Inventory~\cite{hestenes1992force}. See examples of recall, definition probe and concept-inventory items in Appendix \ref{appx:questions}. To facilitate future replication, surveys and other materials are provided in our GitHub repository [link to be added after publication].

\subsection{Procedure}

The study followed a within-subjects design with counterbalanced condition order. Each participant completed two main study blocks: one with the gaze-aware assistant and one with the text-only assistant, using a different reading passage in each block. A third comparison block was then used to directly compare the two analysis types side by side. To maximize comparability across the two main phases, passage order was fixed: participants read the Inflation passage in the first block and the Plate Tectonics passage in the second block, both selected as medium-difficulty texts. For phase 3 passages were randomly sampled across difficulty. See an overview of the procedure in Appendix \ref{appx:appx_accuracy_text}.

At the start of the session, participants provided informed consent, completed a demographics questionnaire, and were fitted with the Aria glasses, followed by eye-tracking calibration. In each of the two main blocks, participants first read a passage and rated their perceived understanding. They then interacted by speech with the assigned assistant. After the interaction, they again rated their understanding and completed the post-task measures, including recall, definition probe, concept-inventory, and exploratory questionnaires. Finally, they were shown the analysis that had informed the assistant in that block and rated its perceived accuracy, confidence in that rating, and personalization.

After completing both main blocks, participants completed a third block designed to compare the analyses more directly. They read an additional passage and were then shown a gaze-based analysis and a text-only analysis side by side in randomized order. For each, they rated perceived accuracy, confidence, and personalization, and indicated which analysis better matched their reading experience. After this, the participants engaged in a 5-minute interview using the explicitation interview technique~\cite{vermersch1994explicitation}, where participants were asked open-ended questions such as "Describe your experience", with follow-up questions probing the users to add further details to their descriptions if details were missing or unclear, for instance, "How did the assistant do X?", "What do you mean when you say Y?".

\section{Analysis}
\label{sec:analysis}
Our main dependent variables were recall, definition probe, and concept-inventory scores of texts across conditions. To assess the LLM's ability to interpret gaze data, our main variables also included participant accuracy ratings of the LLM analysis, their confidence in their rating, and how personalized they found the analysis~(Likert, 1-7). Lastly, in the phase 3 side-by-side comparison, we asked the participants to choose the analysis that they preferred between the eye tracking and text-only based analyses. 

For each dependent variable, we performed within-subject (paired) statistical tests comparing the experimental and control conditions. Specifically, we used paired-sample t-tests to assess mean differences, and Wilcoxon signed-rank tests as a non-parametric alternative when normality was not met. Effect sizes were reported as Cohen’s $d_z$ for paired designs. For binary or ordinal outcomes, we used McNemar’s test or Wilcoxon signed-rank as appropriate. 

Exploratory variables included the change in cognitive load (using NASA-TLX scores) and Likert scales (1-7) for perceived understanding of the text before and after interaction with the assistant, usefulness, AI's ability to understand their cognitive reactions to the content and how engaging they found the AI interaction (using the Human Language Model Interaction Questionnaire~(HLMIQ)). Behavioral metrics included the number of conversational turns and total user words.

We conducted an apriori power analysis using a paired-sample t-test~(two-tailed) in G*Power~($\alpha = .05$, $power = .80$). Assuming a medium effect size~($d = 0.5$, Cohen), the required sample size was estimated at N = 34 participants. Thus, we report results for 36 participants. 

In addition to our main variables, we also conducted an exploratory LLM-as-a-judge analysis of the human-AI conversations in either condition~\cite{zheng2023judging}. This allowed us to investigate how well the model followed instructions as well as which interaction strategies the model used. For the LLM judge, a series of classifier descriptions were put together. The prompts for each classifier can be found in Appendix \ref{appx:llm-judge}. The prompts for the classifiers ranged from sanity checks that the model followed its prompt (e.g.,~'aligned with analysis', 'checked user needs' and 'monitored comprehension'), while others were about the chatbot's ability to use the LLM eye tracking analysis in conversation. We used GPT-5.4 for classification with reasoning effort set to ``None''. 

For the qualitative analysis, interviews were transcribed using Microsoft Word's transcription feature, and a thematic analysis of the transcripts was conducted to find overlapping themes among participants. Inductive coding~\cite{boyatzis1998transforming} was conducted to discover salient themes. Two coders iteratively coded the narratives with themes, discussed and aligned any disagreements, and added emerging themes until reaching theoretical saturation~\cite{creswell2017research}.

\section{Quantitative Results}

\subsection{The Gaze-aware Cognitive Assistant Increased Recall}

Recall performance was higher in the gaze-aware AI condition than in control (control: $88.9\% \pm 15.9\%$, experimental: $96.3\% \pm 10.6\%$; mean paired difference $+7.4$ percentage points, $95\%\ \mathrm{CI}\ [1.3,\ 13.5]$). This effect was significant in both tests ($t$-test: $p = 0.0187$; Wilcoxon: $p = 0.0209$; Cohen’s $d_z = 0.41$).

Definition-probe accuracy showed a smaller, non-significant increase (control: $77.8\% \pm 42.2\%$, experimental: $83.3\% \pm 37.8\%$; mean paired difference $+5.6$ percentage points, $95\%\ \mathrm{CI}\ [-10.5,\ 21.6]$; $t$-test $p = 0.4873$, Wilcoxon $p = 0.4795$, Cohen’s $d_z = 0.12$). Concept-inventory scores were also not significantly different (control: $49.1\% \pm 30.3\%$, experimental: $51.9\% \pm 27.0\%$; mean paired difference $+2.8$ percentage points, $95\%\ \mathrm{CI}\ [-10.3,\ 15.8]$; $t$-test $p = 0.6679$, Wilcoxon $p = 0.7552$, Cohen’s $d_z = 0.07$).

\begin{figure*}
    \centering
    \includegraphics[width=0.99\linewidth]{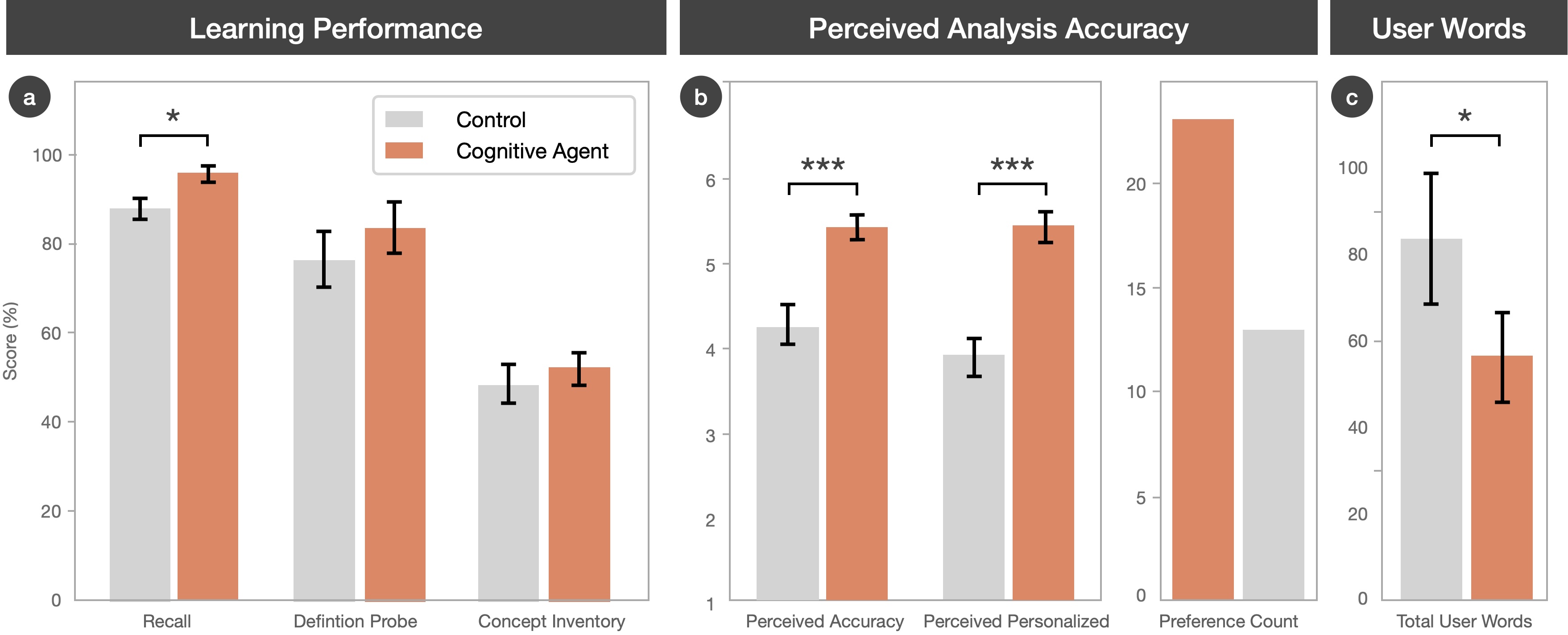}
    \caption{Bar plots of results for main variables across conditions for (a) learning performance, (b) LLM analysis ratings, and (c) the number of words uttered per interactions by each user, with standard errors. * $p<.05$, ** $p<.01$, *** $p<.001$.}
    \label{fig:mainresults}
    \Description{Three sets of bar charts comparing control and gaze-aware AI assistant conditions. 
(a) Learning performance: recall and definition probe scores are higher with the gaze-aware AI assistant, while concept inventory scores are slightly higher but closer across conditions. 
(b) Understanding: perceived understanding ratings are similar between conditions. 
(c) Perceived analysis accuracy: participants rated the gaze-aware AI assistant higher for perceived accuracy and personalization, and more participants expressed a preference for the gaze-aware AI assistant over control. 
Error bars indicate variability.}

\end{figure*}

Taken together, these results indicate a selective learning benefit of gaze-informed support: participants showed a reliable improvement in recall, whereas gains on definition-probe and concept-inventory measures were positive in direction but not statistically conclusive. In contrast, self-reported change in understanding remained similar across conditions (see Fig.~\ref{fig:mainresults}, B ``Understanding"), suggesting that objective performance gains were not always mirrored by subjective learning judgments.

\subsection{Eye Tracking Analyses Were Perceived as More Accurate and Personalized}

The gaze-aware analysis used by the assistant was rated as more accurate and more personalized than analysis for the control condition (see Fig.~\ref{fig:mainresults}, C). Accuracy ratings increased from $4.31 \pm 1.58$ in control to $5.50 \pm 1.06$ in the gaze-aware condition (mean difference $= 1.19$, $95\% \mathrm{CI} [0.60, 1.79]$; paired $t$-test $p = 0.00027$, Wilcoxon $p = 0.00044$, phase 1 and 2 ratings). Personalization ratings likewise increased from $3.94 \pm 1.62$ to $5.50 \pm 1.18$ (mean difference $= 1.56$, $95\%\ \mathrm{CI} [0.99, 2.12]$; paired $t$-test $p < 0.00001$, Wilcoxon $p = 0.00005$, phase 1 and 2 ratings). In the phase-3 comparison, $63.9\%$ of participants preferred the eye-tracking summary (23/36), versus $36.1\%$ preferring the text-only summary (13/36). A secondary analysis of accuracy ratings in phase 3 revealed differences between control and gaze-based analyses to disappear in the more difficult paragraphs (See Appendix \ref{appx:appx_accuracy_text}. This suggests that difficult paragraphs may benefit little from targeted assistance since both control and experimental analyses will likely flag the same things. Overall, these results indicate that grounding feedback in gaze-derived signals substantially improved perceived accuracy and personalization, with a moderate preference advantage in direct side-by-side choice.

\subsection{\textbf{Perceived Effort and Frustration}}

Exploratory measures of workload~(NASA-TLX) and engagement also trended in favor of the gaze-aware AI assistant~(see figure in Appendix \ref{appx:appx_accuracy_text}).  Total user words were significantly lower in experimental compared to control ($83.25 \pm 91.62$ in control to $57.28 \pm 66.57$ in experimental, mean paired difference $=-25.97$ words; Wilcoxon $p=0.0469$; $d_z=-0.29$). Conversation turns showed a similar directional reduction $from (8.39 \pm 3.59) to 7.22 \pm 3.84$; Wilcoxon $p=0.1067$. NASA-TLX outcomes were also directionally lower with the gaze-aware AI assistant, including mental demand ($10.00 \pm 4.81$ vs.\ $8.89 \pm 3.72$), effort ($11.00 \pm 4.58$ vs.\ $10.06 \pm 3.67$), and frustration ($5.17 \pm 4.25$ vs.\ $4.78 \pm 3.96$), but these workload differences were not statistically significant.

\section{Exploratory Interaction Analysis}

To explore assistant behaviors during interactions, we applied an independent LLM-as-judge analysis of the recorded conversations.

\subsection{Both assistants were highly aligned and consistently checked user needs}

One important question is whether the assistant used for interaction in both conditions could accurately ground its interaction in the analyses. Across conditions, both assistants responses were generally rated as highly aligned with the user-needs identified in each analysis ($1.00 \pm 0.00$ experimental vs.\ $0.94 \pm 0.23$ control), and explicit confirmed the users needs with them before giving them explanations ($0.97 \pm 0.17$ experimental vs.\ $1.00 \pm 0.00$ control). The assistant's monitoring of the user's comprehension during the interaction was also scored highly in both conditions ($0.94 \pm 0.23$ experimental vs.\ $0.97 \pm 0.17$ control). These patterns suggest the assistant was able to scaffold assistance when interacting using both control and experimental analyses.

\subsection{Users appeared to steer less, and showed less confusion, with the gaze-aware AI assistant}

The LLM-as-judge analysis also suggested meaningful user-side differences between conditions. 
Participants in the gaze-aware condition were less likely to take the lead in steering the conversation ($0.44 \pm 0.50$) than in control ($0.67 \pm 0.48$; exploratory paired $t=-2.26$, $p=0.030$, $d_z=-0.38$). 
They also showed a lower rate of changing focus ($0.22 \pm 0.42$ vs.\ $0.25 \pm 0.44$) and a lower rate of losing interest ($0.17 \pm 0.38$ vs.\ $0.22 \pm 0.42$), although these differences were small.

Importantly, users were also judged to express confusion less often in the gaze-aware condition ($0.39 \pm 0.49$) than in control ($0.61 \pm 0.49$), with a moderate directional effect (exploratory paired $t=-1.67$, $p=0.103$, $d_z=-0.28$). 
Taken together, these exploratory patterns are consistent with conversations that required less user-side repair and steering in the gaze-aware AI assistant condition and suggest that participants interacting with the gaze-aware AI assistant might have experienced more immediately relevant and helpful content without them needing to steer the dialogue to get the information that they needed. Because these are multiple exploratory comparisons, they should be interpreted as hypothesis-generating rather than confirmatory.

\section{Qualitative Results}
Following the thematic coding method outlined in Section \ref{sec:analysis}, interview transcripts from participants about their experience with the control and cognitive assistant were analyzed with the following themes emerging.

\subsubsection{\textbf{Eye-Tracking Assistance Was Experienced as More Targeted, Adaptive, and Less Repetitive}}
Participants consistently described the gaze-aware cognitive assistant as offering more targeted and less repetitive support than the text-only baseline. Many emphasized that it picked up on the exact concepts they struggled with, which made the interaction feel adaptive and engaging.  

P10 explained: \textit{``it actually got some of the things that I was struggling with. That's what prompted me to actually engage with it and ask more questions versus when it was completely wrong.''} Similarly, P12 noted how the assistant highlighted their difficulties with domain-specific terms: \textit{``it did do a better job of recognizing what I was confused about... it actually got the technical terms I didn’t get right away.''}  

P17 described the benefit of targeted support for difficult vocabulary: \textit{``it was able to kind of walk me through the higher level overview... but I struggled with some of the technical terms and like which one's solid, which one's liquid. It was good at picking out those details and telling them to me in an easier way.''} P8 also remarked on this alignment: \textit{``it actually got some of the things I was like struggling with... that’s what made me want to keep going and ask more.''}  

Other participants echoed that the assistant seemed to “read their mind” by flagging the concepts they had paused on. P3 recalled: \textit{``the second one was like, oh, you might have problems with this as well and that’s exactly what I had just been stuck on.''} P23 similarly highlighted precision: \textit{``the second one was a lot more precise... the first one just assumed I didn’t understand complex words, but the second one actually pointed to where I had reread and lingered. That felt accurate.''}  

Even subtle struggles were picked up. P1 observed: \textit{``the second one actually simplified the parts I was hesitating on, not just repeating the whole passage.''} And P22 noted: \textit{``the second one seemed to propose better ideas and how to better understand the text, almost like it noticed exactly where I had been stuck.''}  

\subsubsection{\textbf{Eye-Tracking Analyses Felt Granular and Personally Relevant While Text-Only Analyses Felt Generic}}

Participants overwhelmingly reported that the gaze-informed analyses felt more precise, individualized, and aligned with their actual reading experience, while the text-only analyses were described as generic or templated. This distinction was reported by many~(P1, P3, P4, P6, P8, P9, P11, P20, P23). The difference was not subtle: participants often remarked that they could “see themselves” in the eye-tracking summaries, whereas the text-only ones felt disconnected from their own struggles.

For example, P4 shared: \textit{``after reading what it had said, I was like, oh, I'm pretty sure that felt exactly like my reading experience.''} Similarly, P23 reflected: \textit{``the second one was a lot more precise... the first one just assumed I didn’t understand complex words, but the second one actually pointed to where I had reread and lingered. That felt accurate.''} P11 highlighted the striking recognition he felt: \textit{``with the quantum computing passage, it picked up on the exact sentence where I was hesitating. I was like, my God, it actually noticed.''}

Some participants emphasized how this specificity fostered trust. P4 admitted: \textit{``At one point I wondered if it was just listing random things, but because it was more specific than the last one, I trusted it.''} Others pointed to the difference in personalization: P8 remarked: \textit{``summary B was more generic, but summary A was aligned with my intentions, it noticed what I was really pausing on.''} P20 similarly explained: \textit{``the second one would just be like what any typical reader might struggle with, but the eye-tracking one was about me specifically.''}

\subsubsection{\textbf{The System Reduced Cognitive Friction and Barriers to Seeking Help}}

Several participants emphasized that the gaze-aware AI assistant lowered friction to asking for help and made clarifications more seamless within the flow of reading. Instead of breaking immersion to look up terms or search on their phone, the assistant provided timely reinforcement that made learning feel more effortless. This was reported by P22, P23, among others.

P23 described this directly: \textit{``so many times I don’t understand something but don’t want to go to my phone and search it up. Here, the friction was lowered, it felt effortless.''}

Others reflected on the system’s potential to make “asking for help” feel natural. By embedding support into the reading process, the assistant removed the social or effortful barriers often associated with stopping to seek clarification. As P23 concluded: \textit{``It’s a better interface to just keep track of everything you’re looking at. So it’s a bit more effortless and more interactive.''}

\subsubsection{\textbf{Misinterpretation of Reading Behaviors}}

Despite reported benefits, participants also pointed out examples where the system had misinterpreted their gaze. Two issues emerged: (1)~rereading or skimming was often misclassified as confusion or skipped parts that they'd need to know, and (2)~internal cognitive processing could not always be captured by surface gaze data. 

Several participants~(P4, P7, P12, P14, P18) reported examples when rereading for confirmation or connecting ideas was treated as a sign of struggle. P7 explained: \textit{``my biggest complaint about it is that sometimes it was interpreting parts that I reread as if I struggled, but actually those were the parts I understood the best.''} Similarly, P4 noted: \textit{``when it said I skimmed stuff, it assumed I needed more help, but I skimmed because I already knew that topic.''} P12 added: \textit{``I was comparing terms [trying to understand their connection], but instead it thought I wanted the definition of each separately.''}

Others highlighted the invisibility of internal cognition. P3 reflected: \textit{``really, in my mind the disconnect was that I didn’t know the connection between the plates and the asthenosphere. I don’t think eye-tracking could ever pick that up.''} P15 also remarked: \textit{``it’s hard to grasp the complexity between the intellectual aspect of reading and then the emotional or personal aspect that the machine can’t see.''}

\section{Discussion}

This work examined whether a gaze-aware AI assistant with access to a user's POV image stream could better identify and address users' likely cognitive needs than a text-only baseline. Compared to the text-only baseline, the gaze-aware assistant produced a significant improvement in recall, was rated as substantially more accurate and more personalized, and was associated with fewer user words during the interaction. Exploratory interaction analyses further suggested that users needed to steer the conversation less and expressed confusion less often in the gaze-aware condition. These quantitative patterns were reinforced by interview data: participants repeatedly described the gaze-aware assistant as more targeted, adaptive, specific, and aligned with the exact concepts or passages they had struggled with, whereas the text-only assistant was often experienced as more generic. Taken together, these results suggest that gaze-aware multimodal assistants can make cognitive assistance feel more personally relevant while also improving some learning-related outcomes.

A key interpretation of these findings is that gaze-based grounding may reduce the burden on users to identify, remember, and articulate where they need help. In the control condition, the assistant had to infer likely difficulties from the text alone, which may have produced plausible but generic support. In contrast, the gaze-aware assistant could draw on behavioral traces of hesitation, rereading, and prolonged attention to propose candidate trouble spots that were more closely tied to each participant’s actual reading experience. This interpretation is supported by multiple parts of our data: participants rated the gaze-based analyses as more accurate and more personalized; in the phase-3 comparison, a majority preferred the gaze-informed analysis; users produced significantly fewer words in the gaze-aware condition; and exploratory interaction analysis suggested less user steering of the conversation. The qualitative data helps explain these patterns further. Participants often said the assistant “got” the exact terms or passages they struggled with, and several described the interaction as more engaging precisely because the system surfaced issues they would otherwise have had to raise themselves. Rather than merely making responses more efficient, gaze-aware support may therefore shift some of the diagnostic work of help-seeking from the user to the system.

The exploratory interaction results further strengthen this interpretation by pointing to a possible interaction-level mechanism. Across conditions, the assistant was generally able to follow the analysis and check users’ needs before explaining, which suggests that the conversational component functioned competently in both settings. The more meaningful differences appeared on the user side: participants in the gaze-aware condition appeared less likely to steer the conversation, and they were judged to express confusion less often. While these exploratory findings should be interpreted cautiously, they align well with both the reduction in user words and the interview reports that the gaze-aware assistant was more precise and less repetitive. Together, these patterns are consistent with conversations that required less repair and less user effort to redirect the assistant toward relevant content. 

Participants’ learning outcomes suggest that the present system’s main benefit may lie more in the \emph{targeting} of assistance than in producing broad gains in learning. The most reliable difference between conditions was on recall, while the definition-probe and concept-inventory measures showed only directional, non-significant improvements, and self-reported understanding changed little. One possible interpretation is that gaze-informed assistance helped participants revisit specific points of local difficulty, such as unfamiliar terminology or passages that had elicited hesitation, without necessarily supporting the kind of generative processing needed for deeper conceptual change. This reading is consistent with participants’ qualitative reports that the gaze-aware assistant often identified the exact terms or sentences they had struggled with. At the same time, the absence of clearer gains on definition and concept-inventory outcomes suggests caution in claiming improved understanding more broadly. This pattern aligns with prior work showing that adaptive and conversational tutors can improve the relevance of support, but that deeper learning and transfer typically depend on instructional strategies that elicit explanation, elaboration, or interactive reasoning from the learner \cite{d2012gaze,dmello2012autotutor,chi1994eliciting,chi2014icap,hausmann2007explaining}. Accordingly, the present findings may be best interpreted as showing that gaze-aware AI assistants can help \emph{find} plausible trouble spots, while stronger evidence is still needed that this form of targeting alone improves higher-order understanding.

Our results also make clear that gaze should not be treated as direct evidence of cognitive need. The qualitative data surfaced a recurring ambiguity: several participants reported that the system sometimes misread rereading, skimming, or cross-checking as confusion, even when those behaviors reflected verification, familiarity, or strategic reading. Others noted that their real difficulty lay not in a specific term they fixated on, but in a deeper conceptual relationship that gaze alone could not reveal. These cases matter theoretically as well as practically. They suggest that gaze is best understood as a probabilistic behavioral cue rather than a ground-truth signal of comprehension breakdown. In our system, this ambiguity may have been amplified by prompting the model to actively search for likely needs, which could encourage over-interpretation of behaviors that were only loosely related to misunderstanding. The same behavioral richness that makes gaze useful for personalization also creates risk of false positives if the system interprets all deviations from smooth reading as evidence of need.


\subsection{Limitations}
The study also has limitations that constrain the strength and scope of our claims. First, the materials were short, and recall performance was high in both conditions, creating some evidence of ceiling effects that may have reduced sensitivity to larger between-condition differences. Second, the control condition was a relatively strong baseline: it was not a generic chatbot, but an assistant explicitly prompted to infer likely user difficulties from the text. This makes the comparison more conservative and isolates the added value of gaze-based information, but it also means the observed differences may underestimate the contrast with ordinary LLM assistance in everyday use. Third, the study used retrospective rather than in-the-moment intervention. Early pilot sessions with real-time interruptions revealed that the assistant frequently
intervened at inopportune moments, disrupting reading flow and frustrating users (even when the underlying inference was correct). This motivated our retrospective design, which reduced interruption risk and made us more able to focus on our research questions around to cognitive need identification and assistance. Future work, should extend our findings to real-time applications and measure their potential benefit.

A further limitation concerns scope of generalization. Our study evaluates the proposed approach in a reading-comprehension setting, where attention targets are spatially anchored and potential support can be localized to words, phrases, or passages. These properties make reading a strong first testbed, but they do not necessarily hold in more open-ended or socially complex activities. Accordingly, our results should be interpreted as support for the feasibility of gaze-grounded assistance in tasks with localizable attentional targets, rather than as evidence that gaze can robustly identify cognitive need across all domains.

\subsection{Design Implications for Gaze-aware AI Assistants}

Our findings suggest several design implications for future gaze-aware and multimodal LLM assistants.

\textbf{Optimizing for targeting vs. optimizing for explanation.} One of the clearest contributions of the gaze-aware assistant was not necessarily richer explanation quality in itself, but better targeting of what to explain. The gains in recall, the strong increases in perceived accuracy and personalization, the reduction in user words, and the exploratory evidence of less steering all point to the value of helping the system identify likely trouble spots more effectively. Future systems should therefore treat need detection and assistance delivery as separate design problems: first identify where support may be needed, then decide how best to respond. For assistance delivery, instructional strategies that promote constructive engagement, such as retrieval prompts, self-explanation, Socratic questioning, or asking users to compare related concepts before providing direct clarification, have been found to more strongly affect learning transfer and could be used to improve assistance delivery \cite{chi2014icap,hausmann2007explaining,d2014confusion}.

\textbf{Behavioral signals are suggestive, not definitive.} Gaze patterns can help surface candidate moments of need, but they should be framed as hypotheses rather than conclusions. Systems should hedge their interpretations, confirm them with the user, and make it easy to reject or redirect an inference when it is wrong. This implication follows directly from our qualitative findings: the same gaze traces that often made support feel highly personalized could also misclassify rereading, skimming, or integrative thinking.

\textbf{Design for low-friction help-seeking without removing user agency.} Participants often valued the system because it lowered the effort required to ask for help and made clarification feel more seamless. This points to a promising role for gaze-aware assistance: reducing the friction of accessing support while still keeping the user in control of whether, when, and how assistance is provided. Rather than interrupting aggressively, future systems may benefit from lightweight prompts, optional follow-up, and interaction designs that let users decide whether to explore a suspected trouble spot further.

\textbf{Support uncertainty and multiple possible readings of behavior.} Because the same gaze pattern can reflect confusion, verification, familiarity, fatigue, or strategic reading, future systems should model uncertainty more explicitly. One practical approach used in this paper is to use substantial hedging in responses (“You seemed to pause here, was that because this term was unclear, or were you connecting it to something else?”) but this still requires an additional step and attention from the user, and will become burdensome, especially for real-time systems which may end up unnecessarily interrupting the user. Future systems should experiment with techniques to improve accuracy such as modeling gaze-behavior semantics for each participant; or collecting and finetuning a gaze-behavior and cognitive needs dataset and large-language model.

\section{Ethics, Consent, and Privacy}

Gaze-aware AI assistants raise privacy and inference risks because eye-tracking and related behavioral signals can reveal sensitive information beyond the immediate task, especially when combined with egocentric video and language-model interpretation \cite{hernandez2021guidelines,abdrabou2025gaze}. Consistent with prior ethics work on affective and behavioral AI, such signals should therefore be treated as probabilistic proxies rather than ground truth about a user’s internal state \cite{hernandez2021guidelines}. In practice, this means systems should clearly communicate what is sensed and inferred, minimize retention of raw multimodal data, and give users meaningful control over when sensing is active and how data are used. Because first-person capture may also incidentally record other people and surrounding environments, privacy protections should extend beyond the primary user alone \cite{alharbi2018can,abdrabou2025gaze}.

\section{Future Directions}

Looking forward, several avenues stand out for deepening and extending this work. Longer and more varied reading materials will be crucial to reveal the selective advantage of gaze-aware systems. Whereas short, uniform passages muted differences between conditions, longer texts would allow generic assistants to default to exhaustive coverage, while gaze-aware AI assistants could concentrate on the specific sections that taxed the reader. Similarly, testing with domain-specific or personally varied materials, for example, programming examples that programmers readily understand but novices do not, could better illustrate how gaze-aware systems adaptively allocate support where it is most needed.

Future iterations of gaze-aware AI assistants should also incorporate additional modalities to reduce misinterpretation. Physiological signals such as pupil dilation, heart rate variability, or EEG might complement gaze data to disambiguate confusion from confirmation or engagement. 

Finally, there is a broader opportunity for these systems to learn over time. Rather than treating each user session in isolation, gaze-aware cognitive assistants could gradually build a model of the user’s prior knowledge, gaze style, and associated difficulties. This might allow the system to distinguish between multiple interpretations and to adjust strategies accordingly. Over repeated interactions, the assistant could learn not only \emph{what} the user struggles with, but also \emph{how} they exhibit it through their behavior.

\section{Conclusion}

This work advances adaptive gaze-aware AI assistants as interactive systems that sense and interpret cognitive needs and adapt assistance accordingly. Across this design space, our study shows that gaze-aware assistance reduces steering, targets help more precisely, and improves recall, while surfacing risks that require explicit governance and meaningful user controls. We are looking towards a future where AI assistants attend to context and physiology with discretion, time assistance to the rhythm of the task, and act as trusted collaborators that lighten cognitive load, deepen understanding, and lead to cognitive growth.



\bibliographystyle{ACM-Reference-Format}
\bibliography{_bib}

\appendix

\section{Prompt Templates and Examples}
\label{appx:prompts}

For reproducibility, we provide the exact prompts used in our system. Each was implemented as a Python function that inserted task-specific variables~(e.g.,~text, gaze wordlist). We group them into three categories: (i)~gaze-based analysis, (ii)~text-only analysis, and (iii)~conversational assistant.

\subsection{Eye Tracking Analysis Prompt}
\begin{lstlisting}
The user is about to engage in a conversation with a helpful, 
personalized chatbot after reading the paragraph:

{paragraph_content}

The user was also looking at the following content with timestamps:
{eye_tracking_wordlist}

# Eye tracking: Were there fixations (words in the same sentence 
looked at repeatedly across seconds)? Regressions (returning to 
previous sentences)? Staring away after a concept? Skipping any 
content (sentences in the paragraph not present in the wordlist)? 
If no significant ones, state so.

# Need help (if any): Based on the most pressing findings, list any 
struggles where the user significantly showed a reaction that might 
require help.

Instructions: Use quantitative observations whenever possible 
(seconds of fixations, regressions, etc.) and align observations 
with the content it was a reaction to.
\end{lstlisting}

\subsection{Paragraph-Only Context Analysis Prompt}
\begin{lstlisting}
The user is about to engage in a conversation with a helpful, 
personalized chatbot after reading the paragraph.

{paragraph}

# Analysis: What is the user most likely to have struggled with?
# Need help (if any): Based on your analysis.
\end{lstlisting}

\subsection{Real-Time Intervention Variant}
\begin{lstlisting}
... [same as Eye Tracking Analysis Prompt] ...

Describe any changes, when they occurred, and what intervention 
(if any) might be needed.

Return the output in JSON with the following fields:
- observations: [...],
- need_help: [...],
- intervention: "Brief opening message the assistant should say 
  if an intervention is needed, otherwise 'none'."
\end{lstlisting}

\subsection{Assistant Prompt}
\begin{lstlisting}
You are a voice assistant that helps people overcome struggles 
when reading paragraphs.

The user just read the following paragraph:
{paragraph}

Below is an analysis of their reading behavior and attention patterns:
{analysis_results}

YOUR INSTRUCTIONS:
- Have a natural, dialogic conversation that helps the user reflect 
  on key ideas and clarify any confusion.
- Tone: Open-ended, Socratic, concise (<20s per turn).
- Be transparent: hedge your inferences ("might", "seems").
- Use the analysis as subtle guidance, starting with what they most 
  likely struggled with. Confirm with the user before moving on.
\end{lstlisting}

\section{LLM Analysis Samples}
\subsection{Example of Generic LLM Analysis}

The paragraph introduces several advanced concepts from quantum mechanics and the philosophy of physics, including:

\begin{itemize}
\item Superdeterminism as a loophole in \textbf{Bell's theorem}
\item The idea of \textbf{local hidden-variable theories}
\item The concept of \textbf{measurement independence}
 \item The correlation between \textbf{measurement settings} and \textbf{hidden variables}
\end{itemize}

The user is most likely to have struggled with:
\begin{enumerate}
    \item \textbf{Understanding Bell's theorem and its assumptions}: The theorem is a foundational but complex result in quantum mechanics, and its assumptions (like measurement independence) may not be familiar to all readers.
    \item \textbf{The meaning of superdeterminism}: The idea that measurement settings and hidden variables could be correlated (thus violating measurement independence) is counterintuitive and philosophically challenging.
    \item \textbf{How superdeterminism allows for local hidden-variable theories to reproduce quantum predictions}: This is a subtle point that requires understanding both the mathematics and the philosophical implications.
    \item \textbf{Terminology}: Terms like "local causality," "hidden variables," "measurement independence," and "Bell inequalities" may be unfamiliar or confusing.
\end{enumerate}

\textbf{Need help (if any)}:
Yes, the user would likely benefit from help, especially with:
\begin{itemize}
    \item A simple explanation of Bell's theorem and its assumptions.
    \item What measurement independence means and why it's important.
    \item How superdeterminism changes the usual interpretation.
    \item An example or analogy to clarify the correlation between measurement settings and hidden variables.
\end{itemize}

\textbf{Summary}: 
The user is likely to need help understanding the technical terms and the conceptual leap involved in superdeterminism, particularly how it relates to Bell's theorem and measurement independence."

\subsection{Example of Eye Tracking LLM Analysis}

\textbf{Fixations (repeated looks within the same idea/sentence)}
\begin{itemize}
    \item Superdeterminism (heading and opening sentence): ~6 looks at 10s, 47s (twice), 48s (twice), 55s.
    \item theorem/theorem. (Bell's theorem references across sentences): frequent returns at 12s, 33s, 49s, 52s, 53s (twice), 62s, 77s.
    \item correlated (correlations with measurement settings): concentrated looks at 43s (twice), 53s, 54s, 68s.
    \item variables/determine/relation near measurement independence: variables at 18s, 42s, 94s; determine at 38s, 93s; relation at 37s, 92s; independence at 95s.
    \item models/hidden-variable: models at 27s, 30s, 83s, 85s; hidden \\ /hidden-variable at 78s, 79s, 24s.
\end{itemize}

\textbf{Regressions (going back to earlier content)}:
\begin{itemize}
    \item After an initial pass through the opening (10s), gaze returned to earlier terms around 47s (Superdeterminism, theorem, correlated, all), indicating re-reading of the intro and key claims.'  
    \item Late-session returns to earlier assumptions (62s) and then to the measurement-independence clause (92s).
    \item Off-text glances (staring at a wall analog)
    \item Brief looks to the monitor bezel/hardware at 46s and again at 61s (~1 second each), away from any on-screen word.
\end{itemize}

\textbf{Skipping content}
\begin{itemize}
    \item No clear skipping. Every sentence of the paragraph appears to have at least one word fixated. The user also looked at the on-screen instruction line (Please read "When you have finished" at 44s.
    \item Overall reading behavior
    \item Typical for dense technical text: initial pass, targeted re-reads of key assumptions and definitions, and a final focus on measurement independence. No signs of major difficulty.
\end{itemize}

\textbf{Need help (if any):}
\begin{itemize}
    \item Clarifying measurement independence (last focal point at 92s): It means the choice of measurement settings is statistically independent of the hidden variables that influence outcomes. Superdeterminism denies this independence by positing built-in correlations between settings and hidden variables.
    \item Connecting local causality with Bell-inequality violation under superdeterminism: Because settings and hidden variables are correlated, the usual Bell assumptions don't all hold, so a local hidden-variable model can match quantum predictions without signaling.
    \item Toy models and what they show: If helpful, I can give a brief, concrete example of a superdeterministic toy model and what assumptions it uses.
    \end{itemize}

Would you like a quick plain-language summary of the paragraph, a short diagram of the Bell assumptions (including measurement independence), or an example toy model?"

\section{Examples of Questions}
\label{appx:questions}

\begin{table*}[t]
\caption{Samples of multiple-choice assessment items by inventory type (Recall, Definition Probe, Concept Inventory) for the main study topics. Correct options are marked with an asterisk (*). Full items can be found in [GitHub repository link anonymized for review]}
\label{tab:sample_inventory_items}
\centering
\scriptsize
\setlength{\tabcolsep}{3pt}
\renewcommand{\arraystretch}{1.1}
\begin{tabular}{p{1.8cm}p{1.6cm}p{3.0cm}p{2.5cm}p{2.5cm}p{2.5cm}p{2.5cm}}
\toprule
Item Type & Topic & Question Stem & A & B & C & D \\
\midrule

Recall & Inflation & Inflation refers to the rate at which ... & wages fall & money supply shrinks & prices rise* & debt grows \\
Recall & Plate Tectonics & The layer beneath the lithosphere is called the ... & lower crust & asthenosphere* & outer core & biosphere \\

Definition Probe & Inflation & An inflation hawk is ... & trader buying gold in recessions & citizen anti-food-tax lobby & policymaker prioritizing low inflation even over growth* & energy market index \\
Definition Probe & Plate Tectonics & In geology, mafic rocks are ... & rich in Mg and Fe (dark)* & silica-rich and light-colored & frozen seawater crystals & formed only by asteroid impacts \\

Concept Inventory & Inflation & Cost-push inflation arises chiefly when ... & demand outpaces supply & input costs (e.g., wages/oil) rise* & taxes are cut & exports surge \\
Concept Inventory & Plate Tectonics & Plate motion is primarily driven by ... & Earth rotation & mantle convection and slab pull* & lunar tides & magnetic storms \\

\bottomrule
\end{tabular}
\end{table*}

\section{LLM-as-a-Judge - Classifier descriptions}
\label{appx:llm-judge}

\begin{lstlisting}
behaviors_nested = {
    "needs_addressed": {
        "description": "For each difficulty flagged in the eye-tracking analysis under 'Need help (if any)', check whether the assistant addressed it in the conversation. Return a JSON object with 'needs_identified' (list of needs) and 'needs_addressed' (list of booleans indicating if each need was addressed).",
        "example": "Assistant explained inflation calculation when analysis showed user struggled with measurement details",
        "response_format": {
            "needs_identified": [
                "Understanding inflation measurement (CPI, core vs headline)",
                "Historical context of 1970s inflation crisis",
                "Definition of inflation hawks"
            ],
            "needs_addressed": [True, False, False],
            "total_needs": 3,
            "addressed_count": 1,
            "score": "1/3"
        }
    }
}

behaviors = {
    "aligned_with_analysis": {
        "description": "Did the assistant's first turn align with the analysis?",
        "example": "Assistant referenced the skipped sentence that the analysis flagged.",
        "0": "not_aligned",
        "1": "aligned"
    },

    "asked_guiding_questions": {
        "description": "Did the assistant ask guiding, open-ended questions?",
        "example": "What do you think the author meant by ``systemic risk''?",
        "0": "not_asked",
        "1": "asked"
    },

    "checked_user_needs": {
        "description": "Did the assistant confirm the user's needs before explaining?",
        "example": "It seems like the second sentence might have been tricky, is that the part you'd like to go over?",
        "0": "not_checked",
        "1": "checked"
    },

    "used_hedging": {
        "description": "Did the assistant hedge its inferences with words like 'might' or 'seems'?",
        "example": "It seems you might have skimmed over this part.",
        "0": "no_hedging",
        "1": "hedging_used"
    },

    "was_concise": {
        "description": "Were the assistant's turns concise (<20s spoken)?",
        "example": "Short reflection and question, not a long lecture.",
        "0": "not_concise",
        "1": "concise"
    },

    "monitored_comprehension": {
        "description": "Did the assistant check whether the user understood before moving on?",
        "example": "Does that explanation make sense to you?",
        "0": "not_monitored",
        "1": "monitored"
    },

    "stayed_on_topic": {
        "description": "Did the assistant stay on topic and avoid introducing unrelated content?",
        "example": "Assistant focused on the paragraph content without diverging to other subjects.",
        "0": "not_stayed_on_topic",
        "1": "stayed_on_topic"
    },

    "user_changed_focus": { # user behaviors
        "description": "Did the user change the topic or focus of the conversation away from the identified needs?",
        "example": "User shifted the discussion to a different part of the paragraph, concept or a new topic entirely.",
        "0": "not_changed_focus",
        "1": "changed_focus"
    },

    "user_expressed_confusion": {
        "description": "Did the user express confusion or misunderstanding about the paragraph content?",
        "example": "User said they were confused about a specific term or concept in the paragraph.",
        "0": "not_expressed_confusion",
        "1": "expressed_confusion"
    },

    "user_engagement_with_assistant": {
        "description": "Did the user actively engage with the assistant's questions and prompts?",
        "example": "User responded thoughtfully to the assistant's questions, indicating active engagement.",
        "0": "not_engaged",
        "1": "engaged"
    },

    # what else couldve gone wrong/right?
    "user_reflected_on_content": {
        "description": "Did the user reflect on the paragraph content during the conversation?",
        "example": "User made comments or observations about the paragraph, indicating reflection.",
        "0": "not_reflected",
        "1": "reflected"
    },

    "user_took_lead": {
        "description": "Did the user take the lead in steering the conversation?",
        "example": "User initiated topics or questions related to the paragraph without prompting from the assistant.",
        "0": "not_took_lead",
        "1": "took_lead"
    },

    "user_requested_clarification": {
        "description": "Did the user request clarification on any points during the conversation?",
        "example": "User asked the assistant to clarify or elaborate on certain explanations.",
        "0": "not_requested_clarification",
        "1": "requested_clarification"
    },

    "user_agreeing_with_assistants_identification_of_their_needs": {
        "description": "Did the user agree with the assistant's identification of their needs?",
        "example": "User confirmed that the assistant accurately identified their need for clarification.",
        "0": "not_agreed",
        "1": "agreed"
    },

    "user_disagreeing_with_assistants_identification_of_their_needs": {
        "description": "Did the user disagree with the assistant's identification of their needs?",
        "example": "User stated that the assistant misunderstood their needs or focus.",
        "0": "not_disagreed",
        "1": "disagreed"
    },

    "user_lost_interest": {
        "description": "Did the user lose interest in the conversation?",
        "example": "User gave short, disengaged responses or indicated boredom.",
        "0": "not_lost_interest",
        "1": "lost_interest"
    },

    "user_needed_more_help": {
        "description": "Did the user indicate they needed more help than what was provided?",
        "example": "User expressed that they still had questions or confusion after the assistant's explanations.",
        "0": "not_needed_more_help",
        "1": "needed_more_help"
    },

    "user_found_explanations_helpful": {
        "description": "Did the user find the assistant's explanations helpful?",
        "example": "User expressed that the explanations clarified their understanding of the paragraph.",
        "0": "not_found_helpful",
        "1": "found_helpful"
    }
}
\end{lstlisting}

\section{Additional Results}
\label{appx:appx_accuracy_text}

\begin{figure*}
    \centering
    \includegraphics[width=0.99\linewidth]{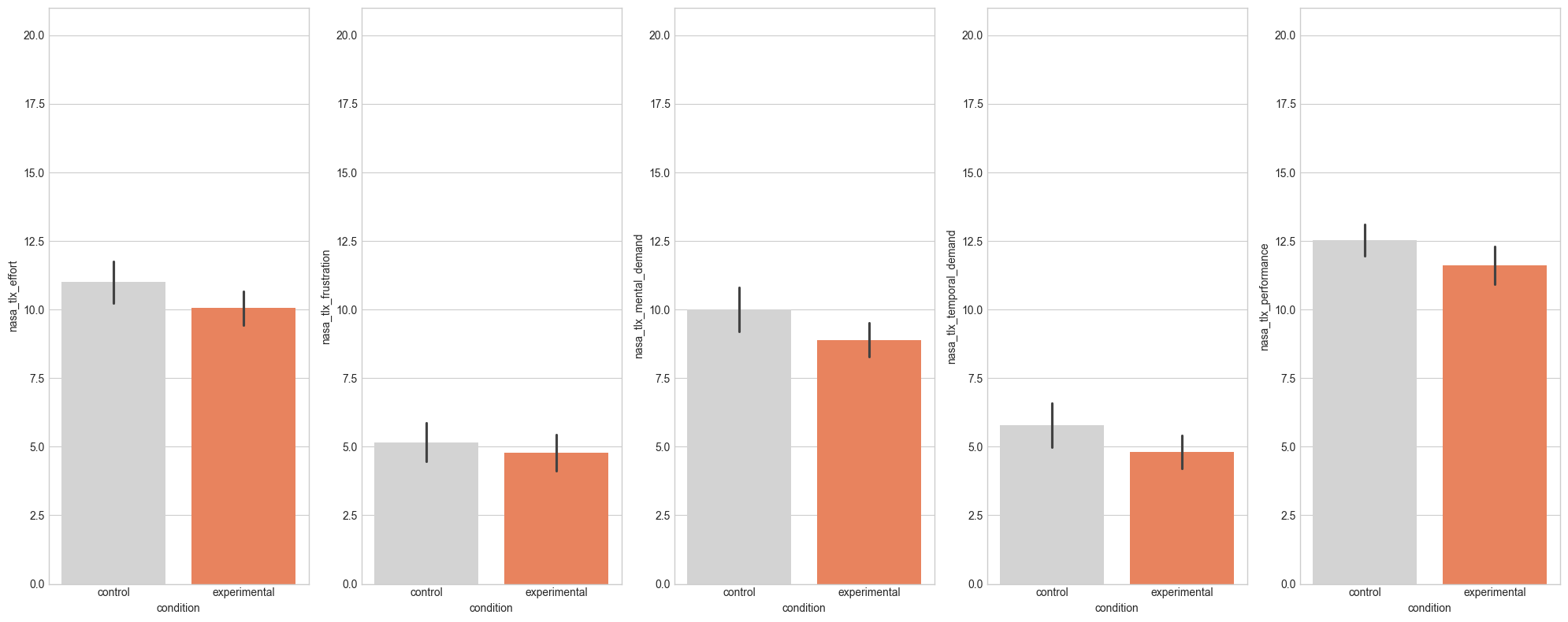}
    \caption{Exploratory results showing NASA TLX across conditions, with standard errors.}
    \label{fig:nasa_tlx}
    \Description{NASA TLX workload ratings: the gaze-aware AI assistant shows lower scores than control for effort, frustration, mental demand, and temporal demand, with performance rated about equal. 
Error bars represent variability.}

\end{figure*}

\begin{figure*}
    \centering
    \includegraphics[width=0.99\linewidth]{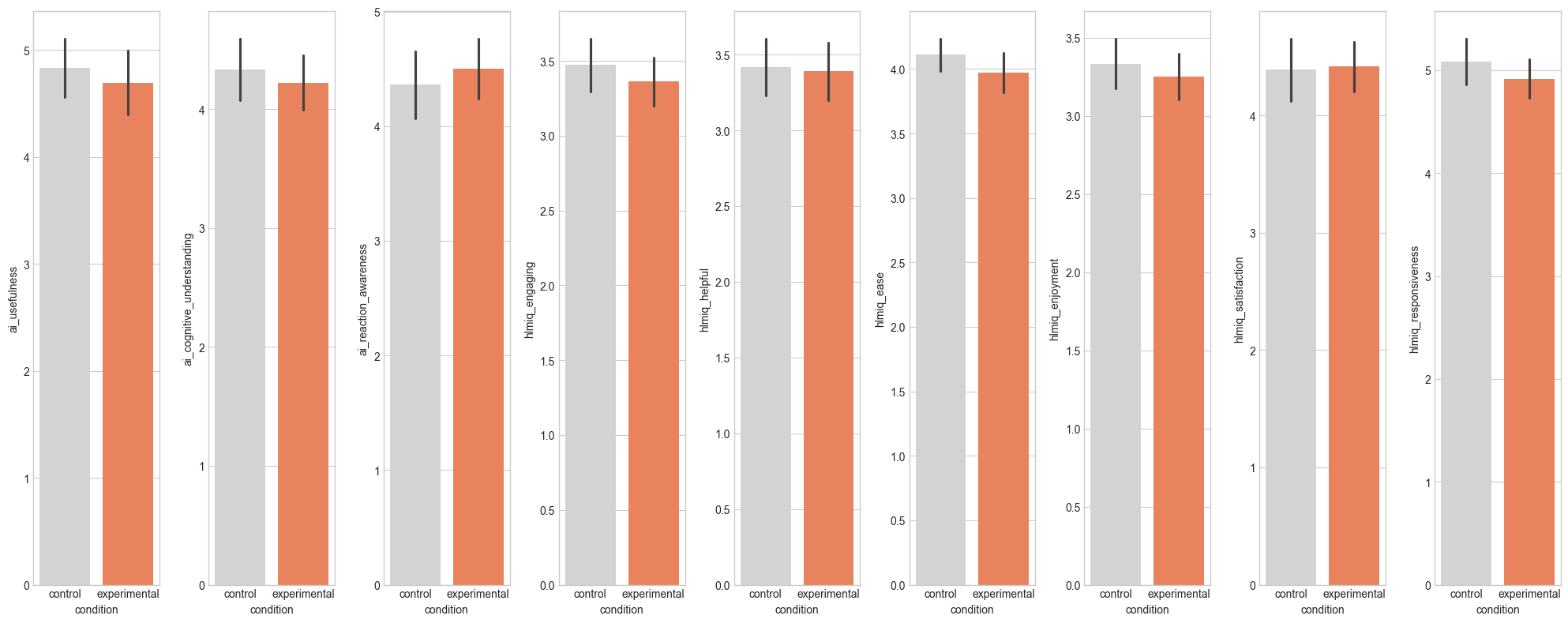}
    \caption{Exploratory results for sanity check questions and HLMIQ responses.}
    \label{fig:user-experience}
    \Description{A series of eight bar charts comparing control (gray) and experimental (orange) conditions on subjective ratings, including perceived usefulness, cognitive understanding, reaction awareness, timing, helpfulness, ease, enjoyment, satisfaction, and responsiveness. Ratings are generally similar across conditions, with small variations: the experimental condition is slightly higher on some dimensions (e.g., reaction awareness, satisfaction) and slightly lower on others. Error bars represent standard errors.}

\end{figure*}

\begin{figure*}
    \centering
    \includegraphics[width=0.99\linewidth]{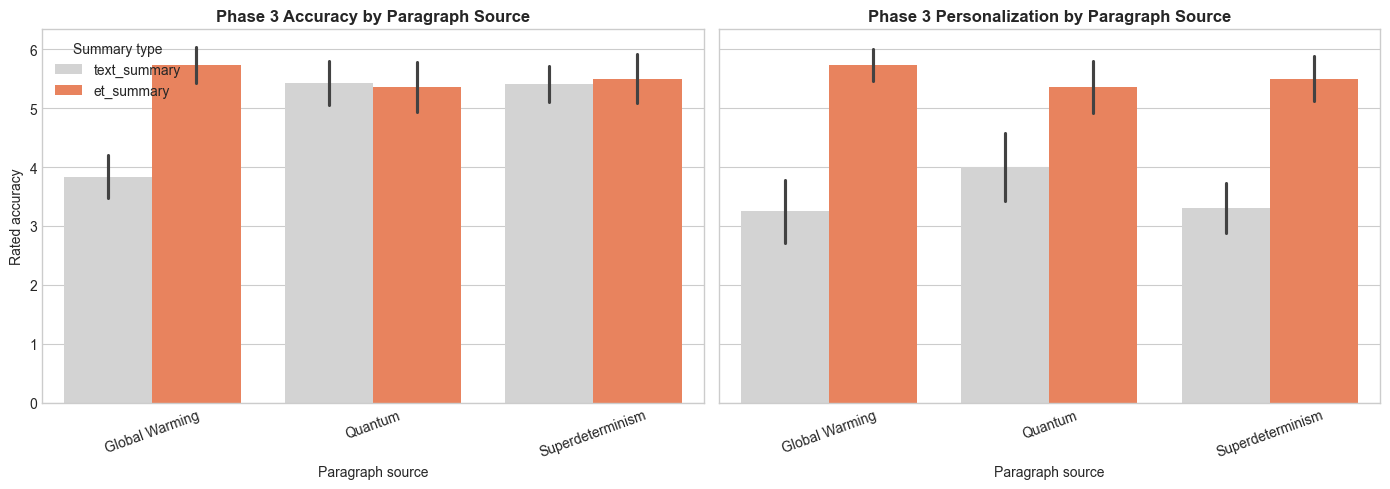}
    \caption{Rated accuracy of AI analysis in phase 3 between control and experimental differed based on text difficulty, with the easier texts showing bigger differences between conditions.}
\end{figure*}

\begin{figure*}[ht]
    \centering
    \includegraphics[width=0.99\linewidth]{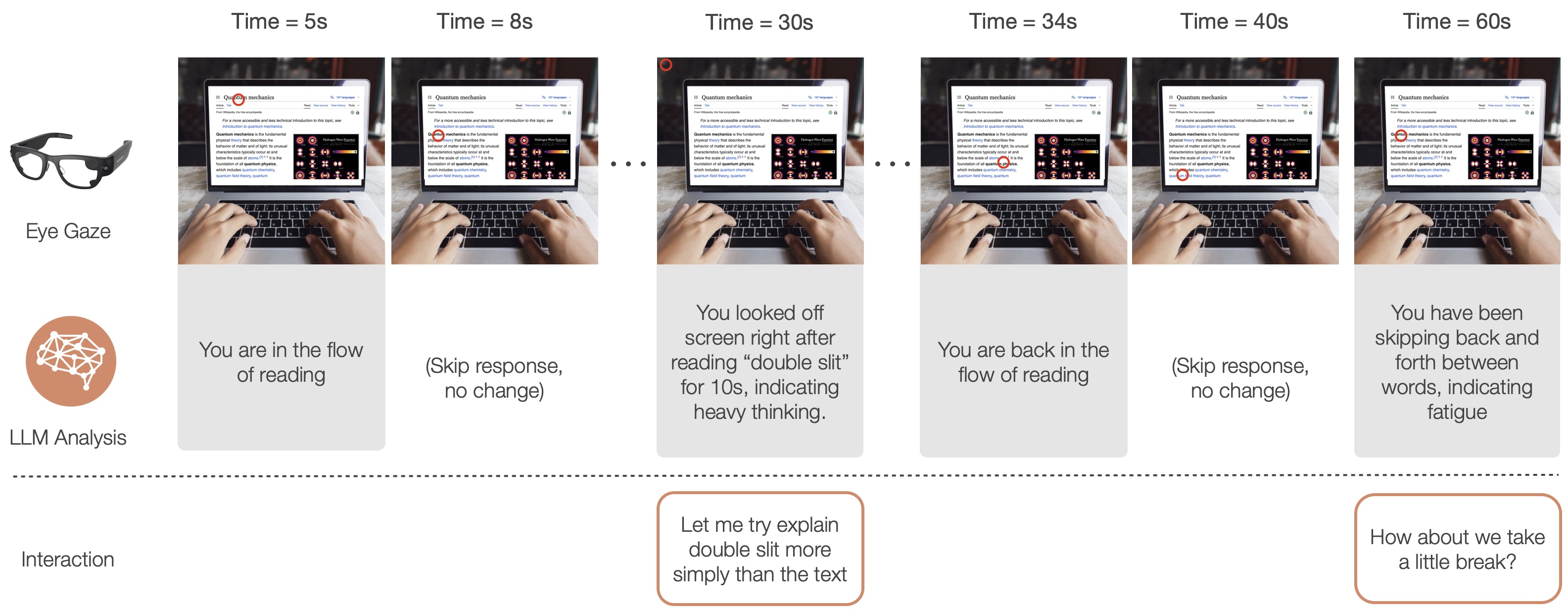}
    \caption{Conceptual overview of system user experience with AI interpretations of gaze behavior. Top row: What is captured with the eye tracking and front-facing cameras. Middle: LLM analysis about the user's current cognitive-emotional reactions to the content. Bottom: Suggested AI interventions.}
    \label{fig:system-user-experience}
    \Description{Timeline of eye-gaze based reading analysis with AI feedback. 
At 5 seconds, the system detects the user is in the flow of reading; at 8 seconds, no change is made. 
At 30 seconds, the user looks away after reading “double slit,” interpreted as heavy thinking. 
At 34 seconds, the user is back in the flow of reading. 
At 40 seconds, no change is made. 
At 60 seconds, the user skips back and forth between words, suggesting fatigue. 
The AI assistant responds contextually: offering to explain “double slit” more simply and later suggesting a break.}

\end{figure*}

\begin{figure*}
    \centering
    \includegraphics[width=0.99\linewidth]{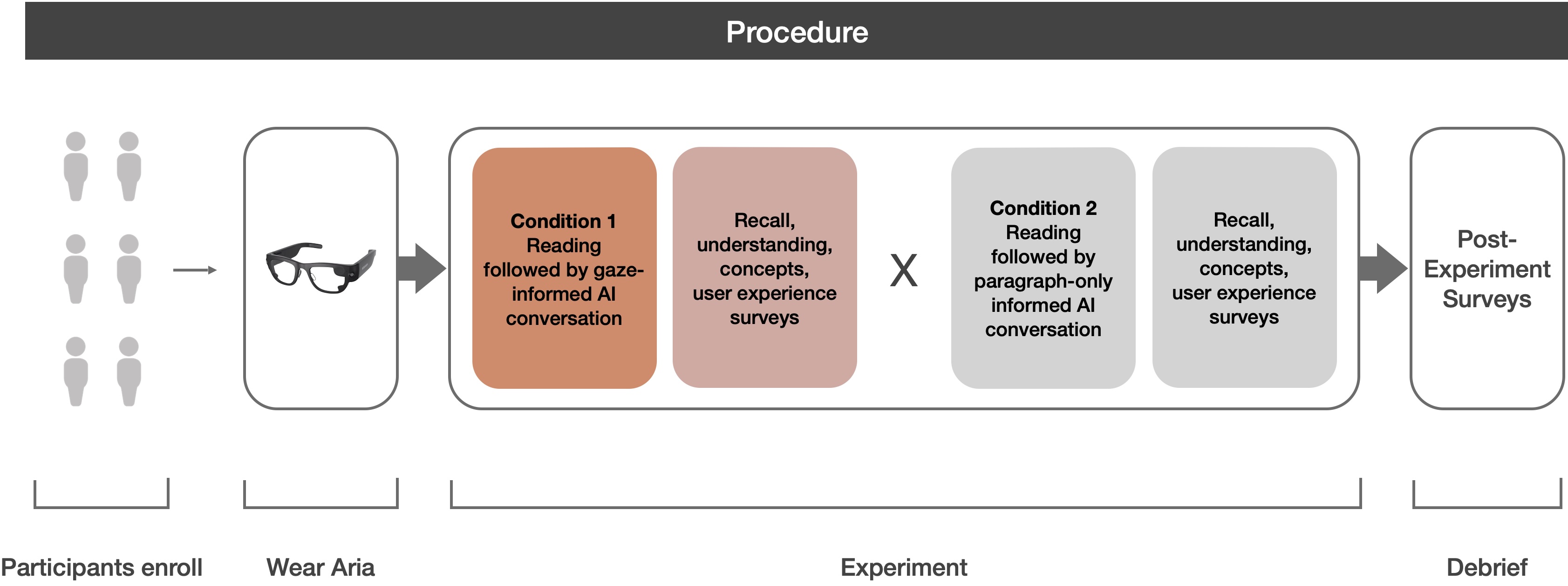}
    \caption{Overview of study procedure and timing across phases.}
    \Description{Flow diagram of the experimental procedure. Participants enroll and wear Aria glasses. 
    In Condition 1, they read a passage followed by a gaze-informed AI conversation, then complete recall, understanding, concept, and user experience surveys. 
    In Condition 2, they read another passage followed by a text-only informed AI conversation, then complete the same surveys. 
    Finally, all participants complete post-experiment surveys during the debrief.}
\end{figure*}

\end{document}